\documentstyle[preprint,aps]{revtex}
\begin{document}
	\input{epsf}
\draft
\title{ 
Truncated-cone SPM tips for surface-molecules interaction studies
}
\author{{\bf R. G. Agostino}, {\bf M. P. De Santo}, 
{\bf G. Liberti}\footnote{Author for correspondence: G. Liberti, 
e-mail: liberti@fermi.fis.unical.it}, {\bf R. Le Pera}\footnote{INFN 
Cosenza},  
{\bf M. Giocondo},\\ {\bf V. Massaro}, {\bf R. Barberi}}
\address{Istituto Nazionale di Fisica della Materia, Unit\`a di 
Cosenza\\
Dipartimento di Fisica, Universit\`a della Calabria, I-87036 Rende 
(CS), Italy}
\date{\today}
\maketitle
\begin{abstract}
A new tecnique for the production of special shaped SPM tips is 
reported. Noble metal coated tips are produced with flat and circular 
tip front. The unusual truncated-cone tip shape is obtained by 
modifying commercial tips. The diameter of the tip front flat area 
can be varied continuously in the range 0.05 $\div$ 1 $\mu$m.
\end{abstract}

\vfill\eject

\section{Introduction}

\begin{figure}
 	\centering
      \mbox{\epsffile[0 0 400 430]{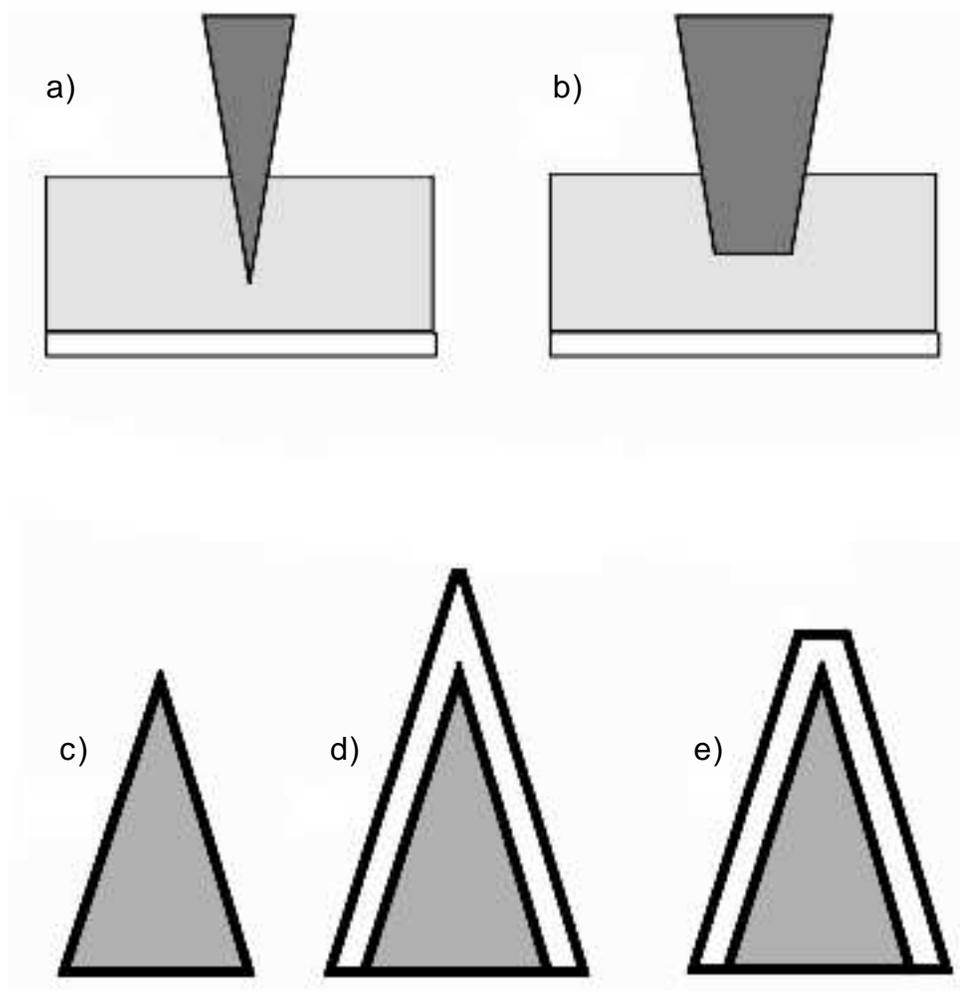}}
 	 \vskip 1.5 cm
\caption{
 \em  In a) is shown a sketch of a typical commercial tip with a sharp end partially 
  immersed in the layer of water that is often present on surfaces in air. b) With new 
  truncated-cone tips, we can have a better insight on forces acting in molecular 
  layers between the tip and the sample. Graphic scheme of the treatment to shape a 
  tip in a truncated cone fashion: c) initially we have the typical sharp tip, d) after we 
  coat it with a thin layer of a noble metal and e) finally we obtain the truncated cone 
  shape by rastering the tip on the atomically flat surface of a Si(111) sample.
 }
\label{fig:1}
\end{figure}

Since the breakthrough of the scanning probe microscopies\cite{Binning1,Binning2} (SPM), 
an important 
step in their evolution was the tip production, whose important features are the 
availability of a good sharpness and the possibility to obtain 
reproducible mechanical 
and electric cha\-ra\-cte\-ri\-stics. It is well established that the use of such tips in a 
Scanning Tunneling Microscope (STM) or in an Atomic Force 
Microscope\cite{Quate} (AFM) 
allows imaging solid surface topography down to the atomic resolution. In 
particular AFM is made in order to be sensitive to forces of the nano-Newton 
order acting between the tip and the surface of a sample.
 The tip can move very close to the sample in vacuum, in air or immersed in a 
liquid\cite{Manne}. Using AFM in air at room temperature, a layer of water is often condensed 
on the surface under investigation. Following the literature, this layer of water is the 
most important source of the attractive force experienced by the tip when it 
approaches the surface (snap-in) and when it is removed from the surface (snap-back). 
The forces acting on a tip when it is immersed in several liquids are 
extensively studied in order to achieve a better understanding of the liquid film 
influence on the imaging process\cite{Weisenhorn1}. In particular, adhesion and friction studies in 
presence of a liquid film on the surface are possible by measuring the torque applied 
to the scanning tip. 
 The ability of the AFM apparatus to reveal forces in the nano-Newton range can be 
used to study the interaction between molecules and surfaces and to test the 
mechanical features of fluids in small volumes. 
 Usually forces between surfaces are measured by means of the so called Surface Force 
Apparatus \cite{Israelachvili}: the involved surfaces are essentially two crossed cylinders and the area 
of contact between them is in the centimeter range. Using AFM with modified 
cantilevers, i.e. attaching micrometer spheres to levers, the contact area is 
reduced to the micrometer range \cite{Biggs}. In both the previous cases the geometry of the 
system is equivalent to a sphere interacting with a flat surface, with 
contact areas always larger than 1 $\mu$m${}^2$. 
 In our opinion, a better insight on molecular forces acting in molecular layers of 
fluids on solid surfaces can be reached using tips having suitable 
shapes and/or decreasing the contact areas down the sub-micrometric scale. 
An appropriate shape could be flat surface parallel to a 
solid surface (Fig.\ref{fig:1}). This kind of tip could be used to investigate the 
elastic properties of micrometric objects, besides modulating the 
interaction force, varying also the size of the contact area 
between the involved surfaces.\par
A more complete knowledge of the elastic properties of the materials 
at small scale has a great importance for instance in the study of liquid crystals (LC). 
The LC macroscopic behaviour is 
highly influenced by the order condition imposed by the surfaces. 
The interaction between the surface and the LC can cause spontaneous orientations 
of the LC molecules that influence the bulk molecules orientation 
\cite{yokoyama}. Physical 
explanations of such processes are not well understood and the existing models are 
essentially macroscopical and phenomenological. 
 The elastic energy stored in a LC sample is expressed as the sum of a term involving 
the energy stored in the volume related to the molecular texture  
in the bulk and a term regarding the interaction  of the molecules with the surface 
(anchoring energy). 
 Measurements on the elastic properties of the LC films can give informations on 
the internal stucture of such films and on the strength of the anchoring conditions. 
Previous observations\cite{Horn,Moreau} were made using a Scanning Force Apparatus on nematic 
and smectic LC, measuring different kinds of forces, each 
reflecting a type of ordering of the LC film near the surfaces and the influence of 
the film thickness on such forces. An AFM with modified tips gives the 
possibility to investigate the same properties locally using surfaces with dimensions 
in the micrometer range, having also the possibility to modify these surfaces in 
order to give different anchoring conditions. It is possible to induce an 
anisotropy on the tip surface or to cover the tip itself with some material to induce 
homeotropic alignment in LC molecules.\par 
 In the present work, we focus our attention on the possibility to produce a new kind 
of tip having a flat surface which can be moved respect to a fixed one, recording the 
interaction forces. This result is obtained by means of a tip with a truncated cone 
shape, with a flat front surface whose area varies from 
$2.5\times 10^{-3}$ to 1 $\mu$m${}^2$. Putting 
this modified tip in a scanning SPM apparatus, it is possible to operate in both 
compression and scanning mode in order to reveal, for instance, the response of a 
fluid to compression, viscosity and friction. Furthermore, it is possible to record the 
influence of an applied voltage between the tip and the fixed surface. 
In the following we describe the method to shape commercial tips in a truncated 
cone fashion. This is achieved by coating\cite{Zhirnov} a Si tip with a thin film of noble metal. A 
similar tip treatment was carried out by H. Andoh et al.\cite{Andoh}; they developed a new tip 
fabrication technique consisting in the evaporation of a thin metal layer on STM 
tips, in order to provide a nanoscale deposition method with different metal species. 
We use, with a different purpose, the same metal deposition technique, adding a 
new feature: the change of the tip morphology, treating it to flatten the front end. 
The first step is performed in vacuum, using the evaporation system described in the 
next paragraph. The final tip treatment is carried out directly in the SPM apparatus 
(Fig.\ref{fig:1}). Force diagrams for standard and flat tips are reported for comparison.

\section{Tip treatment}
\begin{figure}
 	\centering
     \mbox{\epsffile[0 0 400 430]{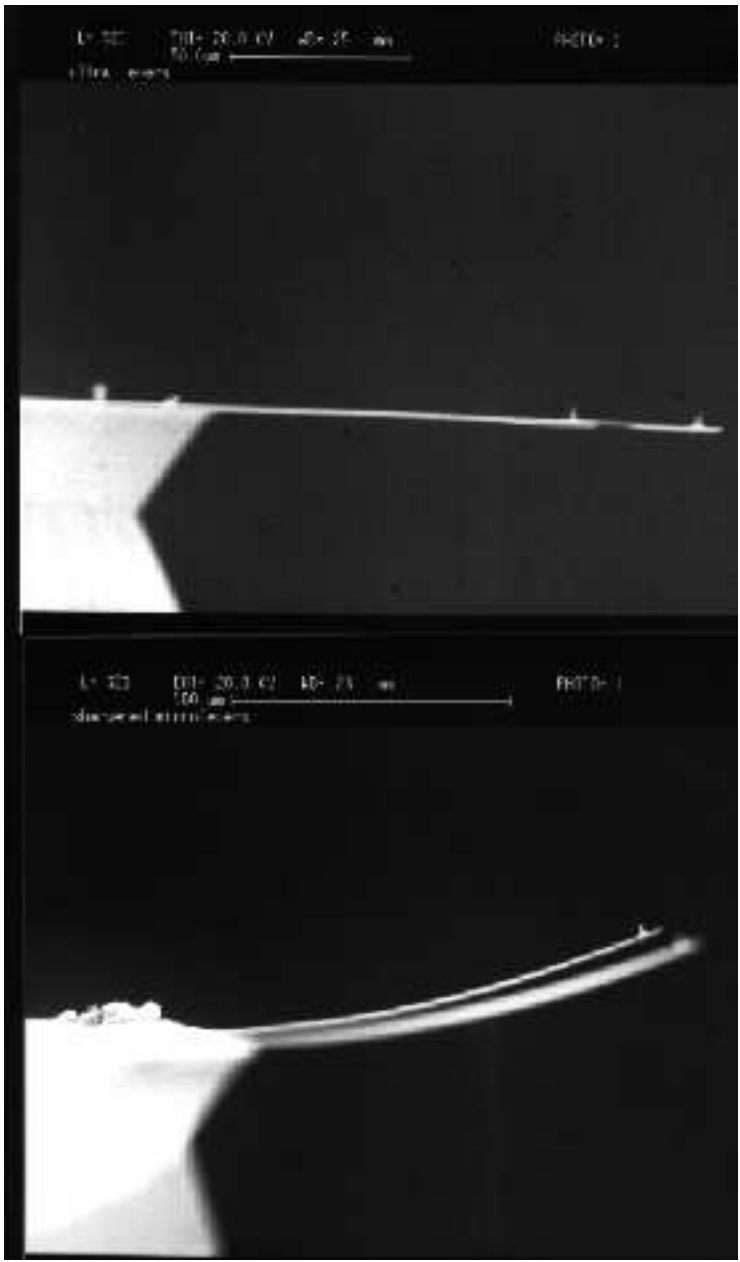}}
 	 \vskip 1.5 cm
\caption{\em  
 a)In this SEM picture is shown a lateral view of PSI cantilevers before the 
  noble metal coating. b)Cantilever deformation due to irradiation, when tips are 
  placed too close (about 10 cm) to the evaporation source.
 }
\label{fig:2}
\end{figure}

A physical vapour deposition (PVD) system with a film thickness control is used for the 
deposition of Ag or Au on standard Si tips. Our coating unit consists of a glass 
chamber (46 cm diameter and 52 cm height) for ultimate vacuum down to 
$10^{-6}$ mbar. The vacuum is achieved in two steps: rough vacuum until 
the pressure of $10^{-2}$ mbar by a mechanical rotary pump and the final high vacuum 
by a turbomolecular pump. 
The coating material is heated in a suitable evaporation oven by Joule effect. 
In our case, the typical working temperature of the evaporation source 
is around 1400 ${}^0C$, where Ag (or Au) sublimates and its vapor deposits on the tip, 
placed in front of the oven. In order to avoid thermal deformation of the cantilevers 
due to irradiation, tips are placed at a minimum distance of 30 cm from the 
evaporation source and a stainless steel shutter with a pin hole is interposed 
between the tip and the oven (Fig.\ref{fig:2}). 
The film thickness is monitored by means of an oscillating quartz balance: the 
oscillating frequency of the quartz crystal changes when a film is deposited on the 
quartz substrate itself. The deposition rate, during the coating process, is fixed at 
about 1 $\AA/s$ and is controlled by means of a rate meter. 
After the coating, the tip is introduced in an AFM apparatus. The AFM used for our 
work is an Autoprobe CP by Park Scientific Instruments with a 10 $\mu$m scanner. 
Typical AFM tips are placed on the edge of a triangular cantilever, whose deflection 
(due to the force between the tip and the sample) is revealed by an optical 
system\cite{Sarid}. 
The optical control is essentially a laser beam, that hits the back of the cantilever 
and is reflected towards a photosensitive detection system (PSPD) .
 Normally, the tip is rastered on the sample surface in order to depict its topography 
through the cantilever deflection revealed by the PSPD signal. 
 We used the scanning apparatus to move a Si tip coated with the Ag (or Au) thin 
film on a Si(111) atomically flat wafer, in order to deform the metal covered tip 
front. The tip is pushed against the Si substrate with a quite high constant force and 
the rastering is performed in {\it constant force mode}. In this case, a feedback circuit 
permits to adjust the cantilever position to keep its deflection constant. 
We have found the following optimized values for the scan parameters to produce 
good truncated-cone tips: rastering on a 4 $\mu$m x 4 $\mu$m area, rastering frequency below 
4 Hz, applied force in the 100 nN range and rastering time from a few minutes to 
1h, when the Ag (or Au) film thickness is a few hundred $\AA$. 
 The tip front radius depends both on the combinations of deposited material 
thickness and  on the applied force and, at present, must be checked for each tip.

\section{Tips characterization}
\begin{figure}
 	\centering
      \mbox{\epsffile[0 0 400 430]{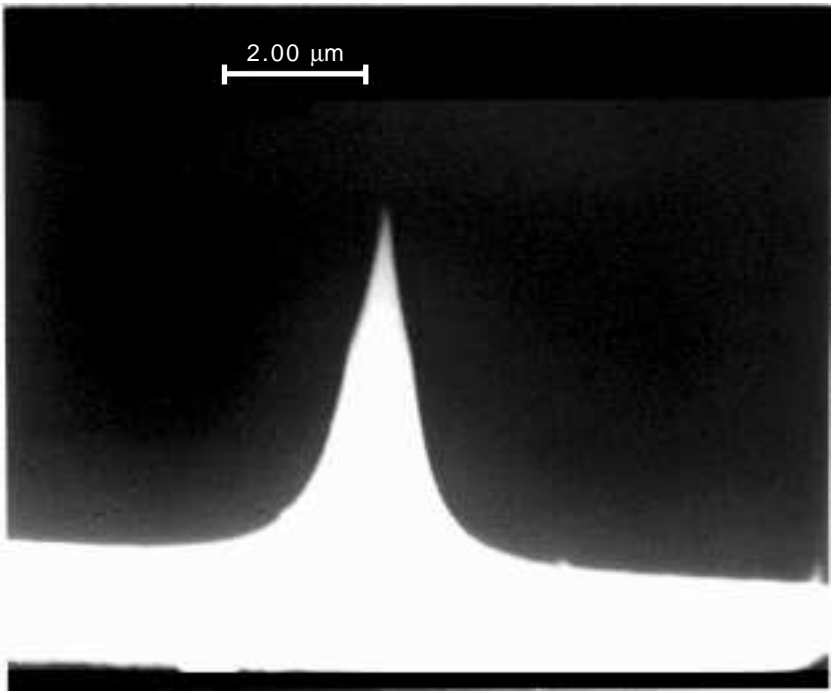}}
 	 \vskip 1.5 cm
\caption{\em  
 SEM picture of a PSI ultralever tip.
}
\label{fig:3}
\end{figure}

\begin{figure}
 	\centering
      \mbox{\epsffile[0 0 400 430]{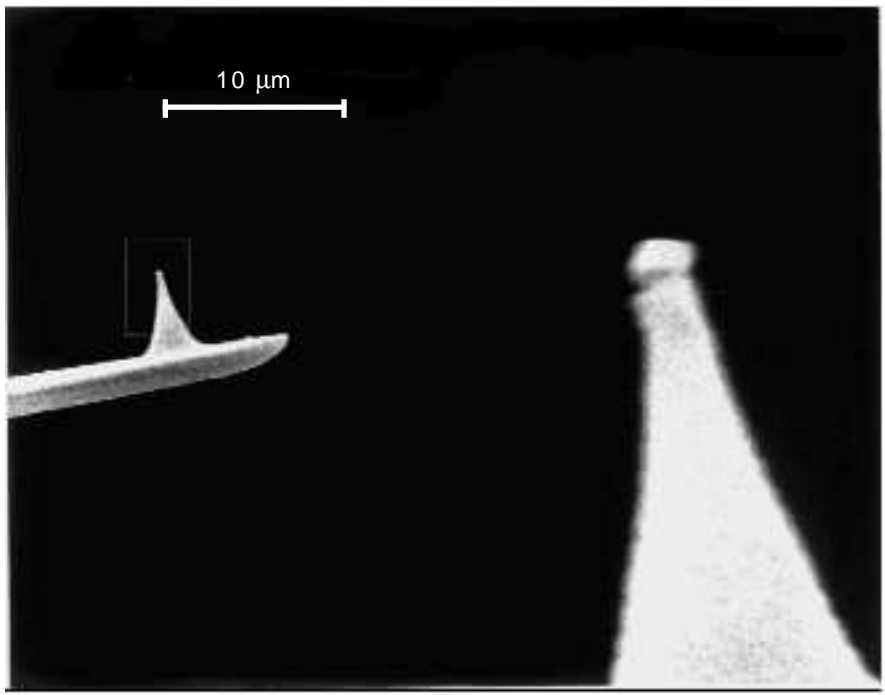}}
 	 \vskip 1.5 cm
\caption{\em  
 SEM picture of an Au coated ultralever tip.
}
\label{fig:4}
\end{figure}

\begin{figure}
 	\centering
      \mbox{\epsffile[0 0 400 430]{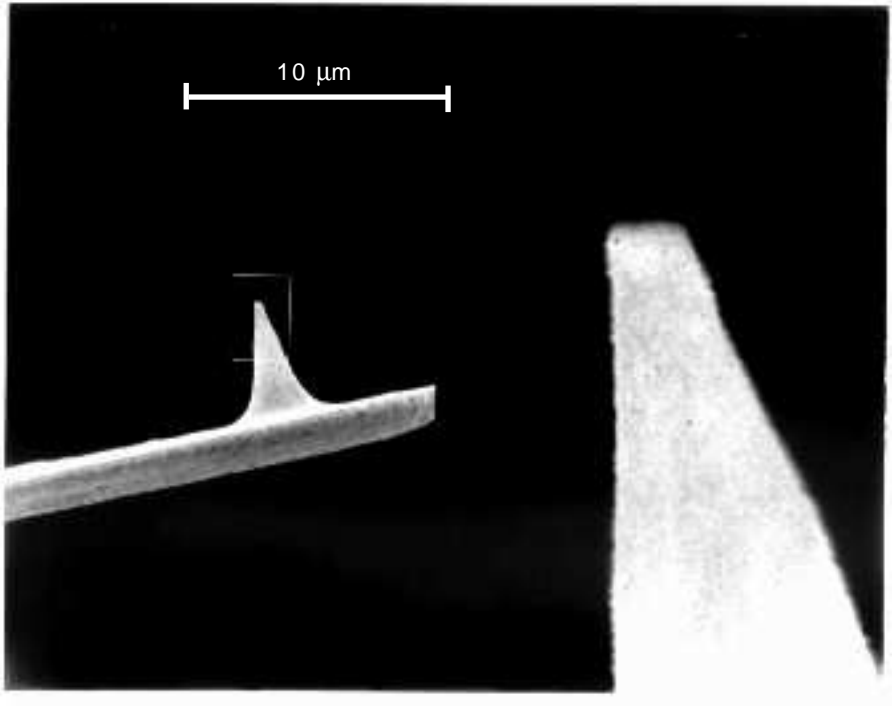}}
 	 \vskip 1.5 cm
\caption{\em  
 Au coated ultralever tip flattened after several scans on a Si(111) substrate 
  in an AFM system.
 }
\label{fig:5}
\end{figure}

 The tip shape is monitored by a scanning electron 
microscope (SEM) at the different steps of treatment. A commercial AFM tip from Park Scientific Instruments is 
imaged in a SEM picture shown in Fig.\ref{fig:3}. It has a high aspect ratio and curvature 
radius lower than 20 nm. The cantilever is attached to a Si chip and its length is 
about 180 $\mu$m.
 After the Au deposition the aspect ratio remains high and the curvature radius is 
now about 0.2 $\mu$m (see Fig.\ref{fig:4}). The deposition is quite uniform and no bumps are 
usually revealed in SEM pictures.
 Fig.\ref{fig:5} shows the SEM picture of an Au coated tip having the flat area diameter of 
about 0.3-0.4 $\mu$m. A further characterization is achieved by 
scanning a freshly Ag coated tip on a calibration grid. Direct information on the tip shape can 
be obtained from the image analysis of grid data. The apparent size 
of determined grid features can be related to the tip radius, that, in the case of a typical 
truncated tip, can be estimated about 700 $\AA$. Assuming that the radius of curvature 
of a commercial tip is about 100 $\AA$, it follows that the Ag coating layer is about 600 
$\AA$ thick, in agreement with the quartz balance reading.

\section{Test measurements}
\begin{figure}
 	\centering
      \mbox{\epsffile[0 0 400 430]{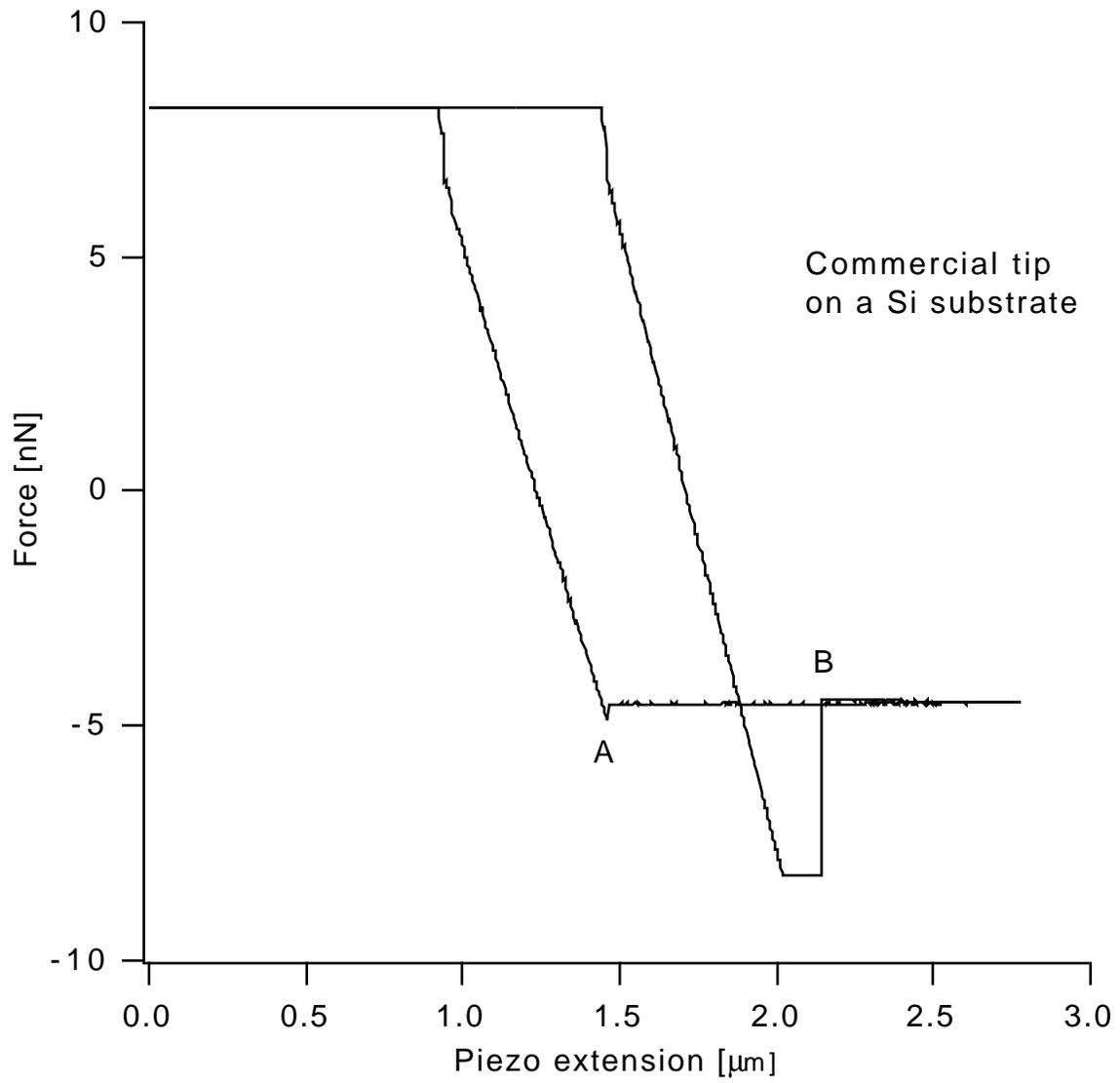}}
 	 \vskip 1.5 cm
\caption{\em  
 Force vs. distance curve of a commercial tip on a Si(111) substrate. A and B 
 are respectively the snap-in point and the snap-back point.
}
\label{fig:6}
\end{figure}

\begin{figure}
 	\centering
      \mbox{\epsffile[0 0 400 430]{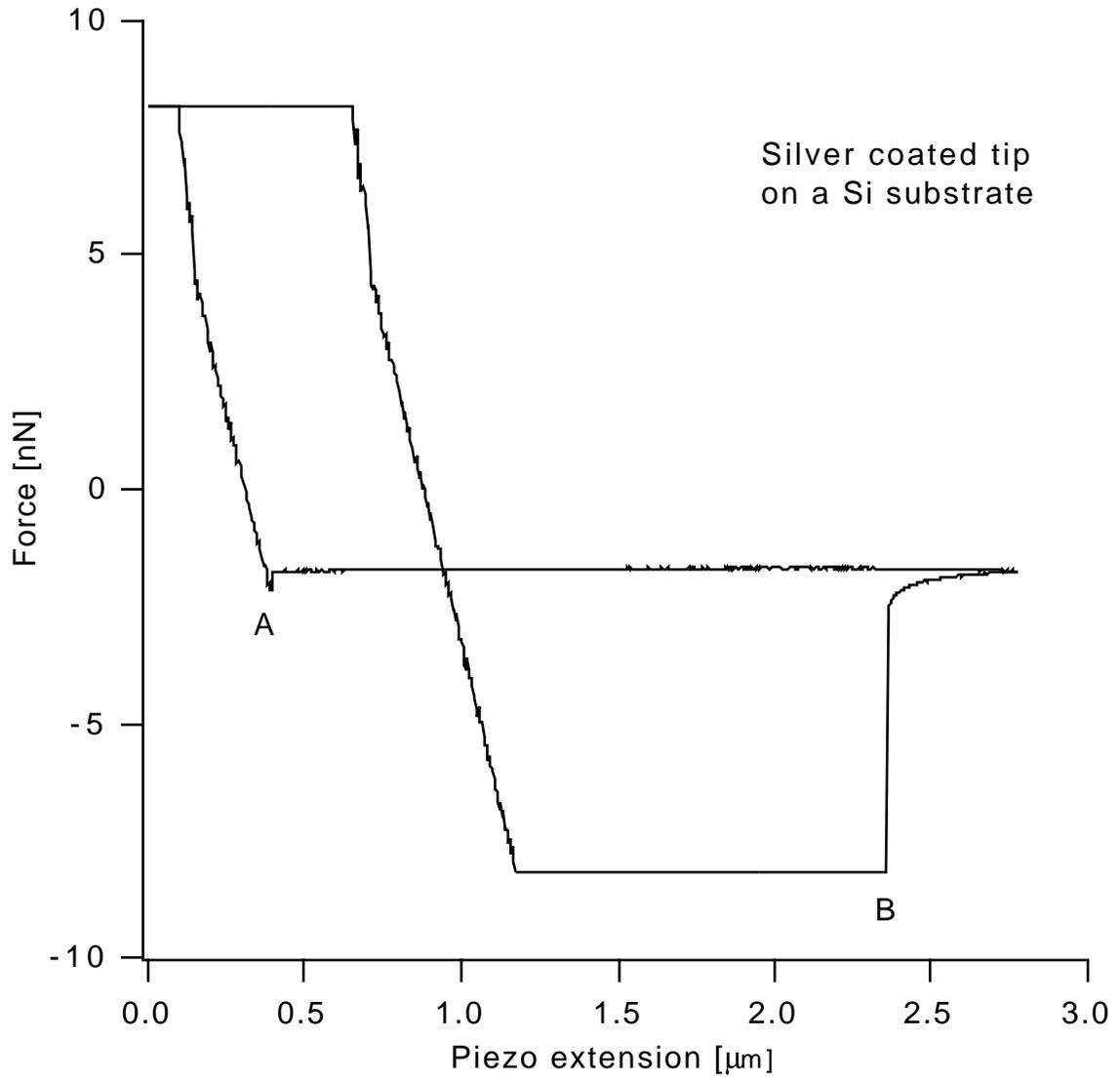}}
 	 \vskip 1.5 cm
\caption{\em 
 Force vs. distance curve of a silver coated tip on a Si(111) substrate with A 
 and B respectively the snap-in and snap-back points. ItÕs worth nothing that the 
  attraction mainly due to capillary forces are bigger than that shown in the previous 
 picture.
}
\label{fig:7}
\end{figure}

In order to compare the different behaviour of a truncated cone tip with a 
sharp one, we acquired a set of force vs. distance curves on a wafer of atomically flat 
Si(111) surface at room temperature and 50$\%$ of relative 
humidity\cite{Binggeli}. 
Forces vs. distance curves\cite{Weisenhorn2,Burnham} are plots of the vertical forces 
between the tip and the 
sample as the z-scanner extends and retracts in a fixed x-y point. These curves can 
be acquired controlling various parameters such as the range of extension of the 
piezo tube (from 0 to 5.564 $\mu$m), the number of acquired points (from 
5 to 2048) and the sweep time (from 0.1 to 600 seconds). Data from 
consecutive scan can be averaged to generate a curve with a lower 
noise-to-signal ratio. The curve  in Fig.\ref{fig:6} is taken with a commercial tip in a 
z range from 0 to 2.782 $\mu$m, using two seconds per sweep and without data averaging. 
In Fig.\ref{fig:7} we show a curve taken with the Ag coated tip with flat end 
in the same conditions as the previous one. 
The two curves show quite similar features. When the scanner starts its path toward 
the surface, the cantilever feels no forces, so it is extended until it experiences the 
van der Waals attractive force (\textit{snap-in point}) and the tip comes in contact with the 
surface. If we continue to extend the scanner, the tip feels the force due to 
the tip-surface repulsion. When the scanner begins to retract, the force follows 
a different path, both in the repulsive part, owing to the nonlinearities of 
the z piezo tube\cite{Hues}, and in the attractive part owing to capillary forces 
due to a thin water layer of the sample surface. 
The scanner retracts until the tip is separated from the water layer 
(\textit{snap-back point}) and a zero-force condition is recovered. 
It's worth noting that the upper and lower limits of the curves are 
cut off, this is due to the saturation of the electronics of the system.
From a comparison between the curves shown in Figs.\ref{fig:6} and \ref{fig:7}, 
we can observe that in the 
retraction path of a flat end tip the snap back feature takes places at 
a greater distance from the surface than in the commercial tip curve. 
The typical elastic constant range of the cantilever is 0.01$\div$0.5 N/m. 
We suppose that the elastic properties of the cantilever itself are unchanged by the coating 
because the thickness of the deposited film which is small with respect to 
the cantilever thickness and to the deposition method, 
that gives a non-compact Ag (or Au) film. Nevertheless the cantilever mass 
increases changing its inertial properties. This does not influence the force 
measurements that can be considered quasi-static.

\section{Interaction forces}
\begin{figure}
  	 \vskip 2.5 cm
	\centering
     \mbox{\epsffile[0 0 400 430]{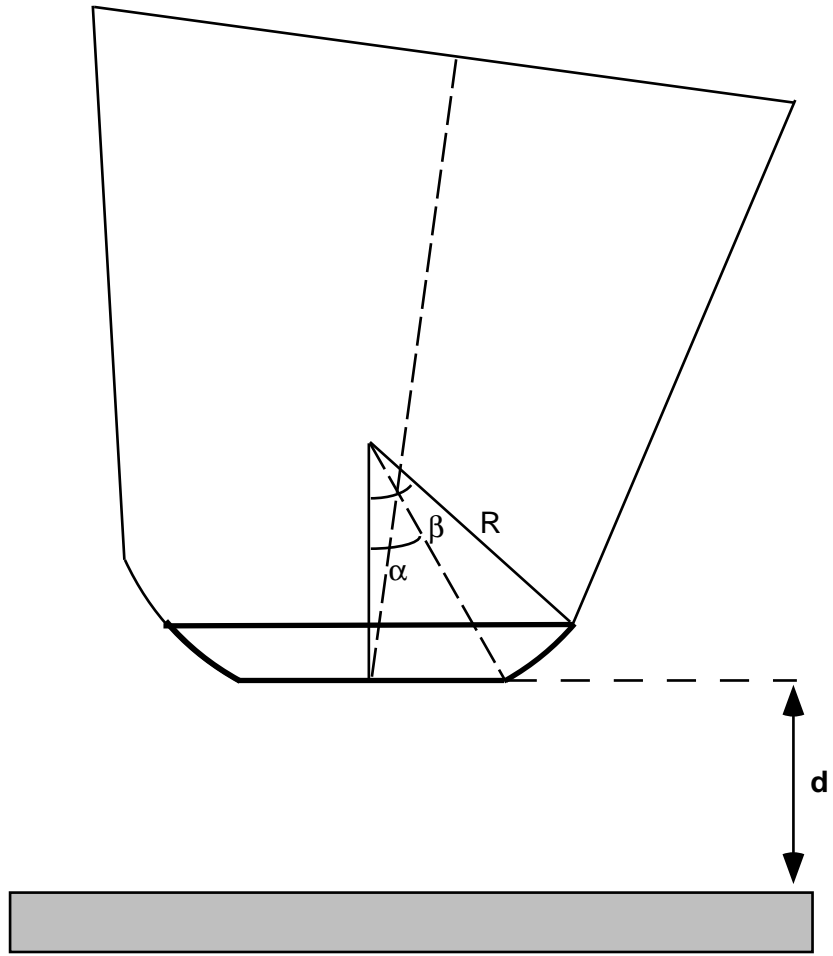}}
 	 \vskip 1.5 cm
\caption{\em  
Model adopted for coated tip-sample system. R is the radius of the 
spherical part of the probe, $\alpha$ and $\beta$ are the angles 
included in the spherical section, $d$ is the probe-sample separation.
}
\label{fig:8}

\end{figure}

As already noted, the snap-in feature is caused by the van der Waals 
interactions between the tip and the sample. We can analyze the London-van der Waals 
interaction forces by modeling the tip as shown in Fig.\ref{fig:8}. It consists of a 
conical section, which surround the entire Si tip (a commercial AFM 
ultralever tip with a cantilever spring constant equal to 0.26 N/m), 
followed by a spherical crown surface and the front flat surface. 
Owing to cantilever deformation due to irradiation and to mechanical 
adjustements during the flattening process, the conical 
section of the tip is not perpendicular to the substrate. It is 
reasonable to assume that the principal contribution to the total 
force on the probe is due to the spherical and flat sections whose 
parameters are the radius of 
the sphere $R$ and the two angles $\alpha$, $\beta$. 
The expression for the total London-van der Waals interaction force between a plane substrate 
and the coated tip can be obtained using the method described in 
\cite{argento} and is given by 
\begin{equation}
	F_{z}(d)=F_{z}^{fs}(d)+F_{z}^{ss}(d)\,,
	\label{eq:1}
\end{equation}¥
where the contribution of the flat surface is
\begin{equation}
	F_{z}^{fs}(d)=-{AR^2\sin^2{\alpha}\over {6 d^3}}\,,
	\label{eq:2}
\end{equation}¥
where $A$ is the non retarded-Hamaker constant\cite{hamaker} and $d$ is 
the separation between the tip and the surface. 
The contribution of the spherical surface is 
\begin{equation}
	F_{z}^{ss}=-{A\over 6}\left\{{R\cos{\alpha}-d\over {d^2}}+
	{d+R(\cos{\alpha}-2\cos{\beta})
	\over{[d+R(\cos{\alpha}-\cos{\beta})]^2}}\right\}\,.
	\label{eq:3}
\end{equation}
By making $\alpha=0$ and $\beta=\pi$ in the above expression we 
obtain the expression for the force-distance relation for a spherical 
tip
\begin{equation}
	F_{z}^{st}=-{2AR^3\over {3 d^2(2R+d)^2}}\,.
	\label{eq:4}
\end{equation}¥
We apply this analysis of the force-distance relations to one 
set of experimental data obtained from a Ag coated truncated-cone probe 
interacting with a flat Ag sample in order to determine the Hamaker 
constant. From the SEM image of the coated tip 
we obtain a tip radius at R = 161.28 nm, and the angles $\alpha=0.378$ rad and 
$\beta=1.292$ rad. The spring constant of the cantilever is fixed at 0.26 N/m.\par
One of major problem in AFM experiments is the determination of the 
point of the zero force and zero separation. We fit each force curve to a 
function of the form
\begin{equation}
	F^m(d)=F_{vdW}(d+d_{0})+B\times(d+d_{0})+C\,,
	\label{eq:15}
\end{equation}¥
where $F^m$ is the measured force, $F_{vdW}$ is the London-van der 
Waals force and the second term represents a linear background ($d_{0}$ 
is the absolute plate separation and $B$ and $C$ are constants). 
The best fit values of $B$, $C$, and $d_{0}$ are 
determined by minimizing the $\chi^2$ and these values are used to 
subtract the systematic errors from the force curve in a region $5\leq 
d\leq 80$ nm and obtain the measured London-van der Waals forces as
\begin{equation}
	F_{vdW}^m(d)=F^m(d)-B d-C\,.
	\label{eq:16}
\end{equation}
The fitted Hamaker constants for 17 different scans ranges from 64 to 
183 zJ and its best estimate is 111.3 $\pm$ 19.4 zJ. The Hamaker 
constant for a gold coated AFM probe on a flat gold surface was 
obtained by Rabinovich and Churaev \cite{Rabinovich} to be in the range 
from 90 to 300 zJ and by Argento and French \cite{argento} to be 126 
zJ.\par
\begin{figure}
 	\centering
      \mbox{\epsffile[0 0 400 430]{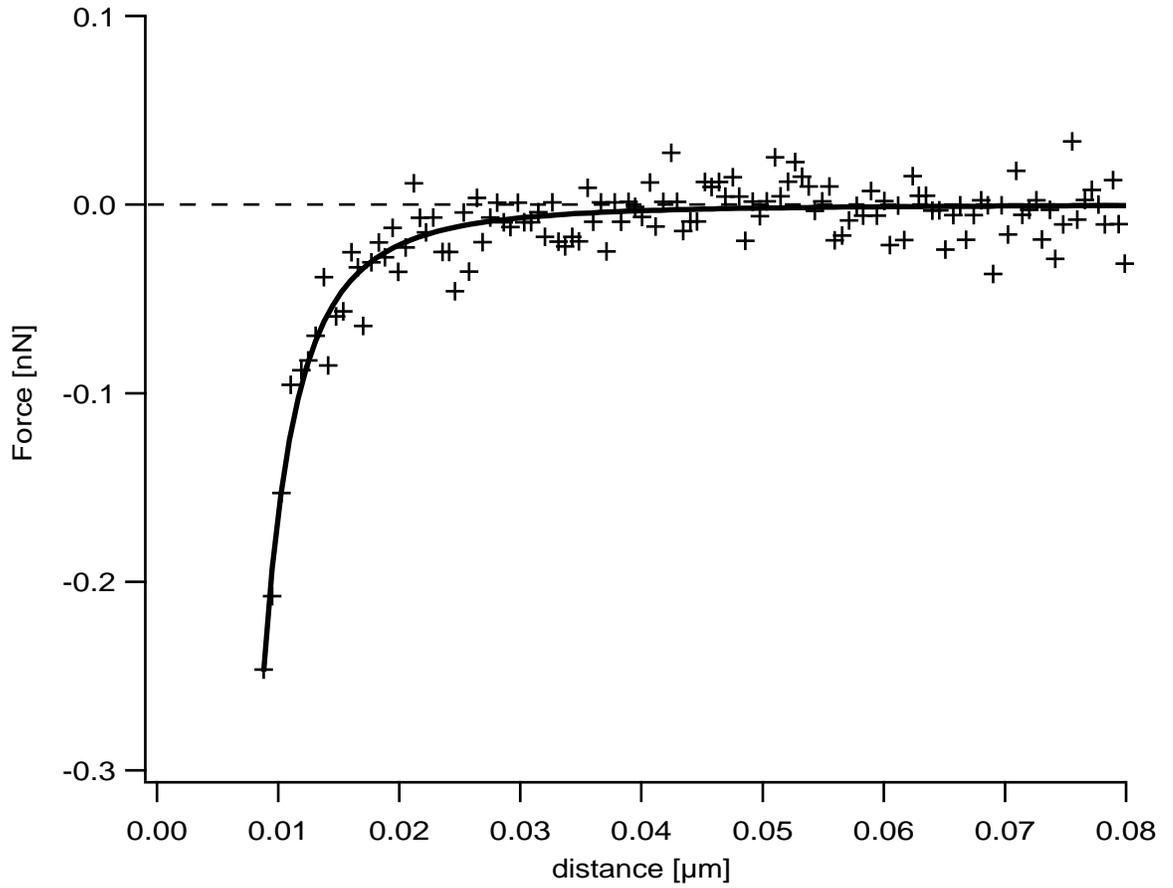}}
 	 \vskip 1.5 cm
\caption{\em 
Experimental data and the resulting fitting for one single scan of a silver coated 
truncated-cone tip on a silver coated substrate. 
The fitted Hamaker constant is 101.46 $\pm$ 15.26 zJ.
}
\label{fig:9}
\end{figure}

 
Fig.\ref{fig:9} shows the experimental data and the fitting obtained with our analysis 
for one single scan. The curve plotted is in good agreement with 
the experimental data.\par
The snap-off features are governed by the capillary forces. This kind 
of interaction is strictly related to the properties of hydrophilicity of the involved 
surfaces and the water layer thickness. We do not give a model for the 
capillary forces in our case because we cannot determine the water 
film characteristics and, thus, we miss a fundamental parameter for 
the force analysis. The commonly used model \cite{Israelachvili} that 
describes the capillary force is not useful because it regards the interaction of a 
curved surface with a flat one, while we have two flat surfaces. 
Intuitively, we expect to have increasing capillary force with the radius of 
the flat tip, but not in proportion respect to the curvature radius of a commercial tip.

\section{Conclusions}
We describe the production and test of truncated cone shaped tips. A noble metal 
film, deposited on commercial tips, is shaped by pushing them with a constant force 
on an atomically flat Si(111) wafer in air. The film thickness and the applied force 
are the main parameters in the front radius control.
We make a first attempt to describe the van der Waals interaction forces between the tip and 
the sample. A comparative test on the capillary forces acting on a normal tip and on 
a flat front tip, shows a clear increase in the latter case. 

\section*{Acknowledgments}
 We thank  Francesco Falco for the kind technical support for the SEM imaging. 
The present work is partially supported by the 1997/99 POP-Regione Calabria 
Fund and INFN-Cosenza.


\newpage 
\listoffigures
\end{document}